\documentclass{article}
\usepackage{fortschritte}
\def\beq{\begin{equation}}                     % 
\def\eeq{\end{equation}}                       %
\def\bea{\begin{eqnarray}}                     %         %
\def\eea{\end{eqnarray}}                       %       % 
\usepackage{amsfonts}
\usepackage{amssymb}
\usepackage{epsfig}
\usepackage{psfig}
\usepackage{psfrag}
\usepackage{dsfont}

\newcommand{\beqn}{\begin{eqnarray}}
\newcommand{\eeqn}{\end{eqnarray}}
\newcommand{\be}{\begin{equation}}
\newcommand{\ee}{\end{equation}}
\newcommand{\non}{\nonumber \\}

\begin {document}                 
\def\titleline{
%%%%%%%%%%%%%%%%%%%%%%%%%%%%%%%%%%%%%%%%%%%%%%
%                                            
% Insert now the text of your title.         
% Make a linebreak in the title with         
%                                            
%            \newtitleline                   
%                                            
%%%%%%%%%%%%%%%%%%%%%%%%%%%%%%%%%%%%%%%%%%%%%%%
\thispagestyle{empty}
\begin{flushright} 
\vspace{-2cm} 
{\small \rm 
MIT-CTP-3454 \\%[-.3cm]
NSF-KITP-03-111 \\[-.3cm]
hep-th/0312172
} 
\end{flushright}
\vspace{1.5cm} 
Brane/Flux Interactions in Orientifolds\footnote{The present article is 
based on the work \cite{Berg:2003ri}.}
%                                             %     %%%%%%%%%%%%%%
%                                             %       %
%%%%%%%%%%%%%%%%%%%%%%%%%%%%%%%%%%%%%%%%%%%%%%%         %
}
\def\email_speaker{
{\tt 
%%%%%%%%%%%%%%%%%%%%%%%%%%%%%%%%%%%%%%%%%%%%%%
%                                                  
% Insert now the e-mail address of the speaker or  
% the author that should get the electronic mail   
% of the publishing house                           
%                                                  
%%%%%%%%%%%%%%%%%%%%%%%%%%%%%%%%%%%%%%%%%%%%%%         %
kors@lns.mit.edu
%                                            %     %%%%%%%%%%%%%%
%                                            %       %       
%%%%%%%%%%%%%%%%%%%%%%%%%%%%%%%%%%%%%%%%%%%%%%         %
}}
\def\authors{
%                                            
%  Insert now the name (names) of the author 
%  (authors).                                
%  In the case of several authors with       
%  different addresses use labels e.g.       
%                                            
%             \1ad  , \2ad  etc.             
%                                            
%  to indicate that a given author has the   
%  address number 1 , 2 , etc.               
%  Signify the person with the above e-mail  
%  address with \sp (behind the address      
%  labels, e.g.                              
%                                            
%            \2ad\sp                         
%                                            
%  when the 2nd author                       
%  happens to be the 'speaker')                                  
%%%%%%%%%%%%%%%%%%%%%%%%%%%%%%%%%%%%%%%%%%%%%%         %
Marcus Berg$^*$, Michael Haack$^*$, and Boris K\"ors$^\dag$ 
%                                            %       %
%%%%%%%%%%%%%%%%%%%%%%%%%%%%%%%%%%%%%%%%%%%%%%         %
}
\def\addresses{
%%%%%%%%%%%%%%%%%%%%%%%%%%%%%%%%%%%%%%%%%%%%%%
%                                            
% List now the address. In the case of       
% several addresses list them by numbers     
% using e.g.                                 
%                                            
%             \1ad , \2ad   etc.             
%                                            
% to numerate address 1 , 2 , etc.           
%                                            
%%%%%%%%%%%%%%%%%%%%%%%%%%%%%%%%%%%%%%%%%%%%%
%
\hbox{
\parbox{7cm}{
\begin{center}
$^*${\it Kavli Institute for Theoretical Physics \\
University of California \\
Santa Barbara, CA 93106-4030, USA \\
}
mberg, mhaack@kitp.ucsb.edu 
\end{center}
}
\hspace{.8cm}
\parbox{8cm}{
\begin{center}                                        %          
$^\dag${\it 
Center for Theoretical Physics \\
Laboratory for Nuclear Science \\
and Department of Physics \\
Massachusetts Institute of Technology \\
Cambridge, MA 02139, USA \\
}
kors@lns.mit.edu
\end{center}
}
}
%%%%%%%%%%%%%%%%%%%%%%%%%%%%%%%%%%%%%%%%%%%%%
\vspace{-1cm}\phantom{}
}
\def\abstracttext{
%%%%%%%%%%%%%%%%%%%%%%%%%%%%%%%%%%%%%%%%%%%%%
%                                             
% Insert now the text of your abstract.       
%                                             
%%%%%%%%%%%%%%%%%%%%%%%%%%%%%%%%%%%%%%%%%%%%%         %
Combining the benefits of D-branes and background fluxes 
in string compactifications opens up the possibility to 
explore phenomenologically interesting brane world models 
with stabilized moduli. However, it is difficult to determine 
interaction effects among open strings and fluxes  
in the effective action. We derive the full bosonic Lagrangian of 
a (spontaneously broken) ${\cal N}=4$ supersymmetric 
model with D3-branes and NSNS and RR 3-form fluxes in an 
orientifold of type IIB, that, without fluxes, would be 
T-dual to type I theory. In the limit where backreaction in form 
of a warp factor is neglected, the effective action can be 
obtained through a procedure that combines 
dimensional reduction and T-duality, and it 
is found to be in agreement with results from gauged supergravity. 
This provides evidence for the consistency of this commonly used 
approximation scheme.
%                                           %       %  
%%%%%%%%%%%%%%%%%%%%%%%%%%%%%%%%%%%%%%%%%%%%%         %
}
\large
\makefront
%%%%%%%%%%%%%%%%%%%%%%%%%%%%%%%%%%%%%%%%%%%%%%%%
%                                              %
%  Insert now the remaining parts of           %
%  your article.                               %
%                                              %
%%%%%%%%%%%%%%%%%%%%%%%%%%%%%%%%%%%%%%%%%%%%%%%%

\setcounter{page}{0}
\vspace{-.5cm}
\section{Introduction}

It has become evident that fluxes, 
i.e.\ vacuum expectation values for certain higher rank tensor fields, 
can provide a step towards solving the moduli problem of string 
compactification (see \cite{Strominger:uh,deWit:1986xg,Polchinski:1995sm}
for the first appearances of fluxes in string theory). 
In the effective four-dimensional field theory, they induce a scalar potential and mass terms for 
many of the otherwise massless scalar fields, that parametrize the size and shape of the compactification space. 
On the other hand, in order to introduce non-abelian gauge symmetries into models within type II or type I string 
theories one has to add D-branes as well. In the absence of fluxes, these wrap calibrated supersymmetric  cycles 
of the background geometry, and the interactions of open and closed string modes are largely dictated by 
supersymmetry and anomaly considerations. Once fluxes are turned on, it is well known that this simple picture may be 
drastically modified, as happens 
in the dielectric effect \cite{Myers:1999ps}. 
The effective action with fluxes is given by a suitably gauged 
version of the original supergravity. On the other hand, since the 
presence of D-branes induces an effective scalar potential 
through the brane tension, the known scalar potential for fluxes 
without branes is modified. We address the issue of flux/brane 
interaction effects in the 
context of a simplified framework \cite{Kachru:2002he,Frey:2002hf}, 
an ${\cal N}=4$ supersymmetric orientifold compactification of type IIB 
with 3-form NSNS and RR fluxes and space-time filling D3-branes, by deriving the full bosonic Lagrangian of 
the effective theory. This will involve the standard gauge theory for the D3-branes and the bulk supergravity 
including the 3-form fluxes, their Chern-Simons (CS) interactions as known for coupled supergravity (SUGRA) and 
super Yang-Mills (YM) systems, 
and finally extra interaction terms that stem from the non-abelian generalization of the effective Born-Infeld (BI) 
brane action, as given in \cite{Myers:1999ps}. The resulting Lagrangian can be compared in great detail to results of gauged 
supergravity \cite{auria}, which confirms the consistency of our approach. \\ 

In order to present the model we are considering \cite{Kachru:2002he,Frey:2002hf}, 
we first have to explain two ingredients of importance. $i)$ The well known no-go theorems (see e.g. \cite{deWit:1986xg}) 
for compactifications of ten-dimensional supergravity 
theories prohibit, under certain circumstances, 
solutions to the ten-dimensional equations of 
motion in type II 
string theory with background fluxes. Following \cite{Giddings:2001yu}
this can be circumvented in orientifold models, which involve (non-dynamical) 
orientifold planes 
(O$p$-planes) with negative energy densities but no perturbative 
degrees of freedom. 
$ii)$ The orientifold model at hand can be understood as a T-dual version of type I string theory compactified on 
a six-torus $\mathbb{T}^6$, T-dualized along all six circles. One may therefore be surprised that it is 
possible to turn on NSNS and RR 3-form fluxes, since neither the original nor the T-dual model possess the 
two types of 2-form potentials as dynamical degrees of freedom. 
We shall explain in which sense the deformation is possible in the T-dual 
model, though type I only has the RR 2-form potential $C_2$ in the spectrum. \\ 

%%%%%%%%%%%%%%%%%%%%%%%%%%%%%%%%%%%%%%%%%%%%%%%%%%%%%%%%%%%%%%%%%%%%%%%%%%%%%%%%%%%%%%%%%%%%%%%%%%%%%%%%%%%%%%%%%%%

\subsection{Orientifolds with fluxes and branes} 

The maybe simplest version of the no-go theorems is obtained from the four-dimensional components of the 
ten-dimensional Einstein equations. If we adopt a general warped ansatz 
for the metric, 
%
%\beqn
$
ds^2_{10} = \Delta^{-1}(x^i) ds^2_4(x^\mu) + \Delta^b(x^i) ds^2_6(x^i),
$
%\eeqn
%
and specify $ds^2_4$ to Minkowski space, 
we find a relation \cite{deWit:1986xg} 
\beqn
g^{\mu\nu} R_{\mu\nu} ~\propto~ 
g^{ij} \nabla_i \left( \Delta^{2(b-1)} \partial_j \ln(\Delta) \right) ~\propto~ 
{\cal T}_{\mu\nu} g^{\mu\nu} - {\cal T}_{ij} g^{ij} \ . 
\eeqn
Any type of matter or energy that satisfies the strong energy condition, as  
background fluxes and D-brane tension do,  
gives a negative definite contribution to the right-hand-side. 
Integrating the total derivative on the left-hand-side then requires fluxes to vanish and D-branes to be absent. 
In other words, there are no compact solutions with warped four-dimensional Minkowski (or de Sitter) vacua in type II 
string theory. There are various ways to circumvent this result, such as 
higher curvature corrections to the ten-dimensional action or the option of breaking maximal four-dimensional symmetry. 
Based on the fact that O-planes formally carry negative tension, the most straightforward way to evade the theorem
appears to be orientifold-, i.e. type I, compactifications or their non-perturbative lift to F-theory. 
In this sense, orientifolds are the unique perturbative string models with a compact internal space, 
that allow simultaneously for fluxes, non-trivial warp factors and D-branes. 
It actually turns out that the structure of the constraints still strictly 
prohibits a positive four-dimensional 
cosmological constant within the framework of static metrics of the type 
we consider, so only Minkowski or anti-de Sitter solutions exist 
\cite{Giddings:2001yu}. There are, however, proposals for getting 
de Sitter backgrounds within string theory, e.g.\ by considering non-perturbative modifications of the four-dimensional 
effective action and anti-D3-branes \cite{Kachru:2003aw}.  \\ 

An important feature of the ten-dimensional solutions is
the warp factor $\Delta(x^i)$ 
in front of the four-dimensional 
metric. On the one hand, it is very attractive in providing an extremely 
rich phenomenology in the spirit 
of Randall-Sundrum brane worlds, 
on the other hand the moduli space for warped compactifications is not 
known in general and a dimensional reduction 
including its effects therefore so far impossible. The usual recipe 
is to neglect the warp factor in the 
reduction to four dimensions. 
Following the arguments given in \cite{Giddings:2001yu}, we assume 
that the warp factor scales like $\Delta(x^i) = 1 + {\cal O}(1/R^2)$ and is 
negligible in the large volume limit $R/\sqrt{\alpha'}\rightarrow \infty$. 
Requiring the overall average radius $R$ to be large compared 
to the string length $\sqrt{\alpha'}$ has the additional benefit 
that it leads to a  
separation of the characteristic mass 
scales, schematically \cite{Kachru:2002he}  
\beqn \label{scales} 
\frac{1}{\sqrt{\alpha'}}\Big|_{\rm string}  \gg \frac{1}{R}\Big|_{\rm KK}  
\gg \frac{{\alpha'}}{R^3} \Big|_{\rm 3-flux}\ .
\eeqn
Summarising, we really reduce on a direct product 
$\mathbb{R}^4 \times {\cal M}^6$ and treat the branes and fluxes 
as a small perturbation on the geometry. 

%%%%%%%%%%%%%%%%%%%%%%%%%%%%%%%%%%%%%%%%%%%%%%%%%%%%%%%%%%%%%%%%%%%%%%%%%%%%%%%%%%%%%%%%%%%%%%%%%%%%%%%%%%%%%%%%%%%%%%%

\subsection{The model: type I$'$ on $\mathbb{T}^6$} 

The definition of type I string theory as an orientifold of type IIB 
implies that its effective action, with 
all open string fields set to zero, is obtained by projecting out all states from the type IIB Lagrangian that are odd 
under the world sheet parity $\Omega$. This leaves a bosonic 
spectrum in ten-dimensional type I consisting of 
$\{ g_{IJ}, \Phi, C_2, A_I^a \}$, the metric, dilaton and RR 2-form, plus the open string vector fields. 
Their coupling to the other fields is actually fixed by supersymmetry and for low energies 
takes the standard form of the type I plus BI Lagrangian. 
As a trivial remark, all fields of type IIB that are odd under $\Omega$ have to vanish identically. \\

We can now imagine to compactify on $\mathbb{T}^6$ and apply six T-dualities to the theory, which effectively means, we project out 
type IIB with the T-dual world sheet parity $\Omega'=\Omega\Theta(-1)^{F_L}$, where $\Theta$ is a reflection of 
all six circles \cite{Kachru:2002he}. 
This leaves us with the spectrum 
$\{ g_{\mu\nu},g_{ij},(B_2)_{i\mu},(C_2)_{i\mu}, \tau, (C_4)_{ijkl}, A_\mu^a, A_i^a \}$, the four-dimensional external 
and six-dimensional internal metric, twelve KK vectors from the NSNS and RR 2-forms, the complex dilaton $\tau = e^{-\Phi} + i C_0$, 
and the open string fields split into internal scalar components and external vector fields. The important observation now is that 
despite the absence of dynamical 2-form potentials 
in the spectrum, the background fields of type IIB do not have to vanish, 
but only be anti-symmetric under $\Theta$, since $f(x^i)=-f(-x^i)$ does not imply $f(x^i)=0$. 
We can then keep the background values for the field strengths $dB_2, dC_2$ of the 
IIB Lagrangian as deformations of the T-dual 
theory. These are 3-form fluxes in IIB, but since 
there are no 2-form potentials in the T-dual theory, they appear as new 
parameters in the Lagrangian, their form and systematics inherited from 
the IIB parent theory. A priori, it does not appear 
guaranteed that this procedure leads to a consistent, 
supersymmetric theory at all. 
One purpose of our investigation is to establish this consistency by matching the action with a gauged supergravity 
Lagrangian. The result will be an explicit and complete bosonic action for the T-dual theory, which we call type I$'$, 
compactified on $\mathbb{T}^6$ with NSNS and RR (i.e. complex) 3-form flux, coupled to D3-brane world volume vectors 
and coordinate scalar fields, 
thus with ${\cal N}=4$ supersymmetry. The steps to perform are: First, 
we T-dualize type I along six circles to deduce the 
T-dual CS couplings to the open string fields from the known form in type I. 
Second, we investigate the non-abelian Born-Infeld 
action in the form given by Myers \cite{Myers:1999ps} 
for extra contributions. Third, we compare with the $\Omega'$-projected 
IIB Lagrangian to deduce the extra deformations of the 
Lagrangian due to 3-form fluxes. 
The objects of greatest interest are the scalar potential that 
arises from the combination of fluxes and branes, and 
the CS action. It is possible to derive the full 
bosonic Lagrangian and match it 
with the gauged ${\cal N}=4$ SUGRA coupled to SYM of \cite{auria}, 
thereby identifying the correct mapping of field variables and parameters \cite{Berg:2003ri}. \\ 

A major drawback of this simplistic model is of course that the ${\cal N}=4$ scenario forbids the appearance of chiral 
matter fields and is thus inappropriate for any phenomenological application. Still we believe the above program to be a useful 
check of the consistency of the approach in general, and hope it will carry over to ${\cal N}=1$ Calabi-Yau compactifications 
(see \cite{Blumenhagen:2003vr} for applications in this direction). \\ 

In \cite{Berg:2003ri} 
it was furthermore shown how the vacua of the effective action correspond to 
solutions of the ten-dimensional equations of 
motion. These solutions involve yet another background for a field strength 
whose potential is projected out of the 
type I$'$ spectrum, the purely internal 
and space-time filling components of the RR 5-form. It is related to 
the warp factor by the Einstein equations and is thus also 
neglected in the reduction. Clearly, it would be interesting 
to generalize the present method for 
constructing deformations of orientifold vacua 
to different T-dual versions of type I, as proposed in the SUGRA framework 
in \cite{Angelantonj:2003rq}. 

%%%%%%%%%%%%%%%%%%%%%%%%%%%%%%%%%%%%%%%%%%%%%%%%%%%%%%%%%%%%%%%%%%%%%%%%%%%%%%%%%%%%%%%%%%%%%%%%%%%%%%%%%%%%%%%%%%%%%%%%%

\section{The effective action with fluxes and open strings} 

To cut short the long and technical story of deriving the effective action, 
we shall be slightly sketchy and 
jump over certain details in the following; 
the full computation can be found in \cite{Berg:2003ri}. 
Our starting point is the ten-dimensional type I Lagrangian 
\beqn \label{typei}
(2\kappa^2_{10}) {\cal L}_{\rm I} ~=~ 
e^{-2 \Phi} \left(R + 4 \partial_\mu \Phi \partial^\mu \Phi \right) - \frac12 |\tilde F_3|^2 
 - \tilde \gamma e^{-\Phi} {\rm tr}\, |{\cal F}|^2\ , 
\eeqn
with $\tilde \gamma = \kappa_{10}^2/g_{10}^2$, 
and otherwise standard definitions for the fields. 
Most important for our purposes is the YM CS correction 
$\omega_3^{\rm YM}$ in the 3-form field strength, 
$\tilde F_3= d C_2 - \tilde\gamma \omega_3^{\rm YM}$. 
After T-duality, this is to be compared to the truncation of 
the type IIB (pseudo-) Lagrangian 
\beqn \label{typeiib} 
(2\kappa^2_{10}) {\cal L}_{\rm IIB} &=& 
e^{-2 \Phi} \Big(R + 4 \partial_\mu \Phi \partial^\mu \Phi
    - \frac12 |H_3|^2 \Big) 
\\ 
&& \hspace{-2cm} 
- \frac12 \Big( |F_1|^2 + |F_3|^2
                 + \frac12 |F_5|^2 \Big)
- \frac{1}{2\cdot 4!\cdot 3!\cdot 3!} \epsilon^{i_0\, \cdots\ i_9} (C_4)_{i_0\, \cdots\ i_3} (dB_2)_{i_4i_5i_6}(dC_2)_{i_7i_8i_9}
\nonumber 
\eeqn
with 
$
F_3 = dC_2 + C_0 H_3,\ H_3 = dB_2,\ F_1 = dC_0
$
and 
\beqn \label{5form}
F_5 & = & dC_4 +
  \frac12 C_2 \wedge dB_2 - \frac12 B_2 \wedge dC_2 \ .
\eeqn
The T-duality of the NSNS part of (\ref{typei}) is completely standard, 
\beqn
G_{ij} \mapsto G^{ij}\ , \quad
A_\mu^i \mapsto B_{\mu i} \ , \quad 
e^{2 \Phi} \mapsto G^{-1} e^{2 \Phi} \ , 
\eeqn
$A_\mu^i$ being the KK vectors from the metric. 

%%%%%%%%%%%%%%%%%%%%%%%%%%%%%%%%%%%%%%%%%%%%%%%%%%%%%%%%%%%%%%%%%%%%%%%%%%%%%%%%%%%%%%%%%%%%%%%%%%%%%%%%%%%%%%%%%%%%%

\subsection{T-duality of RR forms with open strings} 
 
To perform the T-duality of the RR sector \cite{Fukuma:1999jt}, i.e.\ 
of the kinetic term of $\tilde F_3$, it is very helpful to notice that 
the RR kinetic terms of type IIB can be put 
into a manifestly T-duality invariant form. 
To do so, replace the IIB RR and CS Lagrangian, the 
second line of (\ref{typeiib}), by the redundant form 
%
%\beqn \label{sdtypeiib}
%{\cal S}_{\rm RR+CS} ~\rightarrow~
%- \frac{1}{8 \kappa_{10}^2 } 
$\frac14 \left( |F_{1}|^2 + |F_{3}|^2 + |F_{5}|^2 + |F_{7}|^2 + |F_{9}|^2 \right)$, 
%\ , 
%\eeqn 
%
plus impose $* F_1 = F_9,\ 
* F_3 = -F_7,\ 
* F_5 = F_5$ after deriving the equations of motion. The new field strengths 
\beqn \label{defF}
\sum_{p=0}^4 F_{2p+1} = e^{-B_2} \wedge \sum_{q=0}^4 dD_{2q} 
\eeqn
are defined through new RR potentials $D_p(C_q,B_2)$ which can be given 
explicitly. The $F_p$ and $D_p$ 
transform under the T-duality group $O(6,6,\mathbb{R})$ as spinors. 
Now one can apply the element of $O(6,6,\mathbb{R})$ that reflects all 
six circles (in some given order) to $F_3=dC_2$ 
and finds expressions like 
\beqn
(dC_2)^{\{1,2\}}_{\mu ij} & \mapsto & -
\frac{\sqrt{G}}{4!} \epsilon^{ijklmn} (dD_4)^{\{1,4\}}_{\mu klmn} \ , 
\eeqn
where the upper indices $\{ p,q\}$ stand for the form-degree on $\mathbb{R}^4 \times \mathbb{T}^6$, the 
numbers of internal and external indices. This looks qualitatively very much as expected, but the appearance of $D_4$ as 
opposed to $C_4$ is of course crucial here. Keeping track of the KK vectors that appear in the 
contractions in the kinetic terms, one finds that they reproduce the proper terms involving $B_{\mu i}$ in 
(\ref{defF}), and we obtain 
\beqn
F_3 \wedge * F_3 & \mapsto &
F_9^{\{3,6\}} \wedge * F_9^{\{3,6\}} 
+ F_7^{\{2,5\}} \wedge * F_7^{\{2,5\}}
%\non
%&& 
+ F_5^{\{1,4\}} \wedge * F_5^{\{1,4\}}
+ F_3^{\{0,3\}} \wedge * F_3^{\{0,3\}} \ . 
\nonumber
\eeqn
Note that the last term already corresponds to a RR 3-form flux, 
although $D_2^{\{0,2\}}$ is not in the spectrum. 
Adding the CS correction $\omega_3^{\rm YM}$ inside $\tilde F_3$ 
leads to
\beqn 
\hspace{-.5cm}
\tilde F_3 \wedge * \tilde F_3 & \mapsto &
\hat F_9^{\{3,6\}} \wedge * \hat F_9^{\{3,6\}} 
+ \hat F_7^{\{2,5\}} \wedge * \hat F_7^{\{2,5\}} 
%\non 
%&& 
+ \hat F_5^{\{1,4\}} \wedge * \hat F_5^{\{1,4\}}
+ \hat F_3^{\{0,3\}} \wedge * \hat F_3^{\{0,3\}}\ , 
\label{Ftildemap}
\eeqn
where the new CS corrected forms are 
\beqn 
%%&& \hspace{-1cm} 
\hat F_{3+2p}^{\{p,p+3\}} = 
\left[ e^{-B_2} \wedge \sum_{q = 0}^p
\Big( dD_{2+2q} + (-1)^{q(q-1)/2} \gamma 
%            (-1)^{q(q-1)/2} 
\star \omega_3
\Big)^{\{q,q+3\}} \right]^{\{ p,p+3\}} \ ,
\eeqn 
with $\gamma= \sqrt{G}^{-1} \tilde \gamma$, and $\star$ 
denoting six-dimensional 
internal Hodge-duality. 
%Moreover, on the right-hand-side 
%a projection to the $(p,p+3)$-form part is understood . 
Dualizing the kinetic YM term by $A^a_i \mapsto A^{ai}$ 
and splitting ten-dimensional vectors into four-dimensional vectors and scalars produces kinetic 
terms for these, plus 
\beqn 
\sqrt{-g}
  {\rm tr}\, |{\cal F}|^2 & \mapsto &
\frac12 \sqrt{-g_4} 
%\Big( G_{ij} g^{\mu \nu} 
%  \tilde D_\mu A^{ai} \tilde D_\nu A^{aj} 
%%\non
%%&&  
%+~ \frac12 g^{\mu \nu} g^{\rho \sigma} (\tilde {\cal F}^a_{\mu \rho} + H_{i\mu \rho}A^{ai})
%(\tilde {\cal F}^a_{\nu \sigma} + H_{j \nu \sigma}A^{aj})  \non 
%&&  \hspace{2cm} 
%+~ 
G_{ij} G_{kl} f^{abc} f^{ade} A^{bi} A^{ck} A^{dj} A^{el} +\ \cdots \ . 
\eeqn
This is the well known YM contribution to the scalar potential \cite{Witten:1985xb}. 

%%%%%%%%%%%%%%%%%%%%%%%%%%%%%%%%%%%%%%%%%%%%%%%%%%%%%%%%%%%%%%%%%%%%%%%%%%%%%%%%%%%%%%%%%%%%%%%%%%%%%%%%%%%%%%%%

\subsection{Modification due to 3-form fluxes} 

So far we have produced an action that contains RR forms of all degrees. To compare it to the truncated type IIB Lagrangian 
we have to replace some of the forms of unconventional high degree by their Hodge duals. This can be done by a Lagrange 
multiplier procedure replacing $\hat F_p^{\{q,p-q\}}$ by $\hat F_{10-p}^{\{4-q,6+q-p\}}$, 
which is standard to do without background fluxes. 
Before doing this, let us note that the 
formulation in which the BI action for the open string fields is 
usually given in type IIB also contains RR forms of all 
degrees, i.e.\ it provides a redundant version of the 
action, where duality relations must be imposed. 
Thus the redundant form is 
the right framework to investigate any extra modification coming from the 
BI action. Using the notation for 
the non-abelian BI  
action of \cite{Myers:1999ps}, two types of terms
are important for our purposes. The RR forms participate in additional
Wess-Zumino interactions of the schematic form 
\beqn 
\int {\rm tr} \Big( C_4^{\{4,0\}} \ {\rm i}_{A}{\rm i}_{A} \, B_2 \Big)
&\sim& 
\int d^4x\, \sqrt{-g_4} \epsilon^{\mu \nu \rho \sigma}
(C_4)_{\mu \nu \rho \sigma} H_{ijk} \, {\rm tr} (A^i A^j A^k)\ , 
\nonumber 
\\
\int {\rm tr} \Big( {\rm P} \left[ {\rm i}_{A}{\rm i}_{A} \, C_6 \right] \Big)
&\sim&
\int d^4x\, \sqrt{-g_4} \epsilon^{\mu \nu \rho \sigma}
(dC_6)_{\mu \nu \rho \sigma ijk} {\rm tr} (A^i A^j A^k)\ ,
\eeqn
where ${\rm i}_{A}{\rm i}_{A} \, B_2 = A^j A^i B_{ij}$, etc.
The CS interactions of the component $C_4^{\{4,0\}}$, 
non-dynamical in four dimensions, 
are of central importance since its ten-dimensional equation of motion determines the tadpole cancellation 
or RR charge conservation constraint. It is then crucial that the two terms combine into an interaction 
term of the form 
%
%
%\be \label{myersforpres}
%\int d^4x\, \sqrt{-g_4} \epsilon^{\mu \nu \rho \sigma}
%(F_7)_{\mu \nu \rho \sigma ijk} {\rm tr} (A^i A^j A^k) + \ \cdots\
%,
%\ee
$
(F_7)^{\{4,3\}}_{\mu \nu \rho \sigma ijk} {\rm tr} (A^i A^j A^k) + \ \cdots\ ,  
$
and therefore there is no additional contribution to the tadpole 
condition after dualizing $F_7^{\{4,3\}}$ to $F_3^{\{0,3\}}$. 
It then follows that the RR charge effectively carried by the background flux remains unmodified compared 
to the case of trivial open string fields \cite{Giddings:2001yu}, 
\beqn \label{rrtad} 
N_{\rm flux} ~=~ \frac{1}{2\kappa^2_{10}\mu_3} \int F^{\{0,3\}}_3 \wedge H^{\{0,3\}}_3 \ . 
\eeqn 
I.e.\ in contrast to (\ref{Ftildemap}), there is no hat on $F_3$. We come 
back to the physical effects of this at the end of the section.
The second term we need comes from the BI action itself and involves the 
NSNS 3-form flux, 
\beqn \label{nonabten}
\frac{i}{3g_{10}^2 \sqrt{G}} e^{-\Phi} {\rm tr} ( A^{i} A^{j} A^{k} ) (H_3)_{ijk} &=&
- \frac{1}{2\kappa^2_{10}} \frac{1}{3!} \gamma e^{-\Phi} \omega_3^{ijk}(H_3)_{ijk} \ .  
\eeqn
When we add the brane tension to the effective potential, this term 
has to be added as well. Together with 
(\ref{rrtad}), it ensures the positivity of the potential. \\ 

We can now go ahead with the procedure to replace the RR forms of higher degree. 
For the details we again refer to \cite{Berg:2003ri}. 
One imposes the Bianchi-identites of the various field strengths through 
Lagrange multipliers, and then 
integrates out the original forms through their equations of motion, 
leaving the Lagrange multipliers, which roughly are the Hodge-duals 
of the original fields. This allows us to put the action into standard form  
\beqn 
\hat F_9^{\{3,6\}} \wedge * \hat F_9^{\{3,6\}} 
+ \hat F_7^{\{2,5\}} \wedge * \hat F_7^{\{2,5\}}
& \rightarrow & 
%\non 
%&& 
F_1^{\{1,0\}} \wedge * F_1^{\{1,0\}} 
+ F_3^{\{2,1\}} \wedge * F_3^{\{2,1\}} + 2 (2\kappa^2_{10}) {\cal L}_{\rm CS} 
\nonumber
\eeqn
with $F_1, F_3$ as in type IIB. The extra piece in the Lagrangian that is 
generated along the way consists of several CS terms
and reads, after reducing to four dimensions, 
\beqn \label{CSaction}
(2\kappa^2_4) {\cal L}_{\rm CS}  
&=&
-\frac14 \tilde \gamma \epsilon^{\mu \nu \rho \sigma} \Big( 
C_0 \tilde{\cal F}_{\mu \nu}^a \tilde{\cal F}_{\rho \sigma}^a 
%\non 
%&& %\quad\quad %\mbox{} \hspace{3.8cm} 
- 2 \left(F_{j\mu \nu} - C_0 H_{j\mu \nu}\right)
\Big( A^{aj} \tilde{\cal F}_{\rho \sigma}^a
+ \frac12 A^{aj} A^{ai} H_{i\rho \sigma} \Big) \Big)   \non
&&
+ \frac{1}{4\cdot 4!} \epsilon^{\mu\nu\rho\sigma} \epsilon^{ijklmn} C_{ijkl} 
 F_{m\mu\nu} H_{n\rho\sigma}\ . \label{CS}
\eeqn
The first line is the new CS action that couples open string and 
KK vector fields and scalars, while the second line 
is the type IIB bulk CS term subject to projection with $\Omega'$. In 
the presence of fluxes the bulk CS term gets additional contributions, 
not displayed in (\ref{CS}), which can be obtained by demanding 
invariance under gauge-symmetry, see \cite{Berg:2003ri}.
Note that $\tilde {\cal F}_{\mu\nu}$ refers to redefined gauge fields  
$\tilde A^a_\mu = A^a_\mu - A^{ai} B_{\mu i}$. 
We find it a very satisfactory check of the 
methods used that (\ref{CSaction}) matches the expressions given in \cite{auria} 
for the coupled ${\cal N}=4$ SUGRA-SYM 
Lagrangian. \\ 

Finally, we can deduce the additional terms that arise when 
projecting the type IIB Lagrangian (\ref{typeiib}) 
in the simultaneous presence of NSNS and RR 3-form fluxes  
with the modified world sheet parity $\Omega'$. 
We just note two significant examples. The kinetic terms for the axionic 
scalars that descend from the internal RR 4-form $C_4$ in (\ref{5form}) are modified by 
\beqn
&& 
\partial_\mu (C_4)_{ijkl}
- 2 (B_2)_{\mu [i} ( dC_2
+2 \gamma \star \omega_3 )_{jkl]}
+ \gamma (\star \omega_3)_{\mu ijkl} 
~\rightarrow~ \non 
&&
\partial_\mu (C_4)_{ijkl}
- 2 (B_2)_{\mu [i} ( dC_2
+2 \gamma \star \omega_3 )_{jkl]}
+ \gamma  (\star \omega_3)_{\mu ijkl} 
%\non 
%&& \hspace{-1cm}
%\quad\quad\quad\ \ \, \,
+ 2 (C_2)_{\mu [i} (dB_2)_{jkl]} \ ,
%\quad\quad\quad\ \ \, + \tilde\gamma  (\star \omega_3)_{\mu ijkl} \ , 
\eeqn 
i.e. extra St\"uckelberg mass terms appear in $|\hat F_5^{\{1,4\}}|^2$. 
The scalar potential, which was already present through the ``kinetic term'' of $\hat F_3^{\{0,3\}}$, is extended to 
\beqn
&& 
|(dC_2 + \gamma \star \omega_3 )^{\{0,3\}}|^2 ~\rightarrow~ 
%\non
%&&
|(dD_2 + \gamma \star \omega_3 )^{\{0,3\}}|^2 + e^{-2\Phi}|H_3^{\{0,3\}}|^2 = 
|\hat G_3^{\{0,3\}}|^2 \ , 
\eeqn
i.e.\ the (CS corrected) complex IIB 3-form flux $\hat G_3^{\{0,3\}}$ is completed. 
As already mentioned, one also has to add  
the tension of the localized  
sources (D3-branes and O3-planes), including 
the extra non-abelian term (\ref{nonabten}), to the flux energy  
\beqn \label{fullG}
|\hat G_3^{\{0,3\}}|^2 - \frac{1}{18} e^{-\Phi} \epsilon^{ijklmn} (F_3)_{ijk} (H_3)_{lmn} 
 - \frac{4i}{3} \gamma e^{-\Phi} {\rm tr}(A^iA^jA^k) (H_3)_{ijk} ~=~ 
2 |\hat G_3^{\rm ISD}|^2 \ , 
\eeqn
ISD indicating the imaginary-self-dual part of $\hat G_3^{\{0,3\}}$ under internal Hodge duality \cite{Giddings:2001yu}. 
The second term is the contribution of the tension and 
it is crucial therefore that (\ref{rrtad}) did not 
contain $\hat F_3^{\{0,3\}}$. Otherwise (\ref{fullG}) 
would not have come out positive definite. 
This would then have also implied 
that even if $\hat G_3^{\rm ISD}=0$, the typical coupling that drives the 
dielectric effect would not have vanished, contrary to expectation. 

%%%%%%%%%%%%%%%%%%%%%%%%%%%%%%%%%%%%%%%%%%%%%%%%%%%%%%%%%%%%%%%%%%%%%%%%%%%%%%%%%%%%%%%%%%%%%%%%%%%%%%%%%%%%%%

\subsection{Some consequences} 

To compare with the results of gauged supergravity,
we transform to the four-dimensional Einstein-frame, yielding 
the full effective potential in agreement with \cite{auria}, 
\beqn \label{fullpot} 
(2\kappa^2_4) {\cal V}_{\rm eff} = \frac{e^\Phi}{\sqrt{G}} 
|(G_3 + \gamma\star\omega_3)^{\rm ISD}|^2
 + \frac{\tilde\gamma e^\Phi}{2 G}
G_{ij} G_{kl} f^{abc} f^{ade} A^{bi} A^{ck} A^{dj} A^{el} \ . 
%g_{ij} g_{kl} f^{abc} f^{ade} A^{bi} A^{ck} A^{dj} A^{el} \ . 
\eeqn
It is positive definite and of the no-scale type.
It has been stressed in \cite{deAlwis:2003sn} that this is 
consistent with the absence of solutions to 
the ten-dimensional equations of motion with a 
positive four-dimensional cosmological constant, due to the 
fact that the volume modulus does not have a stable minimum.
Indeed the global minima of the potential, its Minkowski vacua, 
are characterized by  
%
%\beqn \label{vacua} 
%{\cal V}_{\rm eff} = 0 ~\Leftrightarrow~ 
$
(G_3 + \gamma\star\omega_3)^{\rm ISD} = % 0 , % \quad 
f^{abc} A^{ib} A^{jc} = 0 . 
$
%\eeqn
%
Since these conditions scale trivially under rescaling of the 
total volume $R^6$, $R$ is a free modulus. This 
also provides us with the opportunity to meet the self-consistency 
requirement (\ref{scales}) 
by simply choosing the free parameter $R$ large in terms of $\sqrt{\alpha'}$. 
On the other hand, one eventually has to break up the no-scale 
structure to stabilize the volume, possibly by higher derivative 
\cite{Becker:2002nn} or non-perturbative \cite{Kachru:2003aw} corrections 
to the effective action. Then (\ref{scales}) has to be met as a 
restriction on the vacuum, 
otherwise the scalar potential (\ref{fullpot}) is not reliable. \\

%%%%%%%%%%%%%%%%%%%%%%%%%%%%%%%%%%%%%%%%%%% 
\begin{center}
{\bf Acknowledgements} 
\end{center}  
B.K.\ is grateful 
to the organizers of the {\it 36th International Symposium Ahrenshoop on 
the Theory of Elementary Particles, 
Recent Developments in String/M-Theory and Field Theory}, who created 
an inspiring atmosphere for a great workshop. 
The work of B.~K.\ 
was supported by the German Science Foundation (DFG) and in part by
funds provided by the U.S. Department of Energy (D.O.E.) under cooperative 
research agreement $\#$DF-FC02-94ER40818. 
The work of M.~B.\ and M.~H.\ was supported in part by I.N.F.N., by the
E.C. RTN programs HPRN-CT-2000-00122 and HPRN-CT-2000-00148, by the
INTAS contract 99-1-590, by the MURST-COFIN contract 2001-025492,
by the NATO contract PST.CLG.978785 and 
by the National Science Foundation under
Grant No. PHY99-07949. Moreover, M.~B. was supported by
a Marie Curie Fellowship, contract \# HPMF-CT-2001-01311 and by the 
Wenner-Gren Foundations, and M.~H. by the German Science Foundation (DFG). 

%%%%%%%%%%%%%%%%%%%%%%

\end{document}